\documentclass[preprintnumbers,amsmath,amssymb,showpacs]{revtex4}

\usepackage{graphicx}
\usepackage{dcolumn}
\usepackage{bm}
\usepackage{graphicx}
\usepackage{booktabs}
\usepackage{CJK}
\usepackage{color} 

\begin{document}

\title{Preparation of $km$-photon concatenated GHZ states for observing distinct quantum effects at macroscopic scale}

\author{ Dong Ding$^{1,2}$}
\author{Fengli Yan $^1$}
\email{flyan@hebtu.edu.cn}
 \author{Ting Gao$^3$}
 \email{gaoting@hebtu.edu.cn}

\affiliation {$^1$ College of Physics Science and Information Engineering, Hebei Normal University, Shijiazhuang 050024, China\\
$^2$Department of Basic Curriculum, North China Institute of Science
and Technology, Beijing 101601, China\\
$^3$College of Mathematics and Information Science, Hebei
Normal University, Shijiazhuang 050024, China\\}

\date{\today}

\begin{abstract}
As a class of multipartite entangled states, the multipartite concatenated GHZ (C-GHZ) states remain superior stability under the influence of decoherence. We propose two scalable experimental realization of the multiphoton C-GHZ states based on the entanglers of multiphoton GHZ state. Given a $km$-photon GHZ state as an input state, if $m$ is odd, one can create a $km$-photon C-GHZ state. Also, generally, we design a scheme to prepare $km$-photon C-GHZ states from $km$ single-photon states by using $k$ entanglers of $m$-photon GHZ state and $k$ $m$-control Toffoli gates.

\end{abstract}

\maketitle

\section{Introduction}
Entanglement is a fundamental quantum mechanical resource that plays a pivotal role in many of the most interesting applications of quantum computation and quantum information \cite{NC2000}. As a central concept, entanglement run counter to the intuition, for example, the famous Schr$\ddot{\text{o}}$dinger-cat gedanken experiment \cite{S1935}. While it predicts quantum mechanics in principle allows superpositions of macroscopic states.
Recently, in the setting of the Hilbert space of $km$ two-level systems, Fr$\ddot{\text{o}}$wis and D$\ddot{\text{u}}$r \cite{FrowisDur2011} study the stability of superpositions of macroscopical quantum states under decoherence. In their Letter the authors introduce a class of quantum states, namely, the concatenated GHZ (C-GHZ) states and discuss a possible experimental realization of the C-GHZ states using trapped ions based on the multipartite M${\o}$lmer-S${\o}$rensen gate \cite{MSgate2000}. The so-called C-GHZ state which is robust against noise and decoherence is defined as
\begin{equation}\label{}
 |\phi_{\texttt{C}}\rangle=\frac{1}{\sqrt{2}}({\left|\texttt{GHZ}_{m}^{+}\right\rangle}^{\otimes k}+{\left|\texttt{GHZ}_{m}^{-}\right\rangle}^{\otimes k}),
\end{equation}
where ${\left|\texttt{GHZ}_{m}^{\pm}\right\rangle}=\frac{1}{\sqrt{2}}({\left|0\right\rangle}^{\otimes m} \pm {\left|1\right\rangle}^{\otimes m})$ are $m$-GHZ states as maximally entangled states in $m$-qubit systems \cite{Gisin1998,Yan2011}
and $k$ is the number of logical qubits, each built of $m$ physical qubits.
Further, through a comparison with different encodings (GHZ encoding, cluster states encoding
and product state encoding), it has been shown that concatenated GHZ encoding
is in fact optimal for the stability of the trace norm of the interference terms \cite{FrowisDur2012}. In other
words, there are superior stability for the C-GHZ states.

Optical quantum systems are prominent candidates for quantum computation and quantum information \cite{Kok2007}, since the photon, the smallest unit of quantum information, is potentially free from decoherence. Because photons do not naturally interact with each other, the key challenge in simulating quantum systems is the construction of two-qubit (multi-qubit) quantum operations, for example, controlled-NOT gate. As a matter of fact, one can induce an effective interaction between photons by making projective measurements with photodetectors. In 2001, Knill \emph{et al} \cite{KLM2001} showed a scheme for efficient quantum computation with linear optics, where the probabilistic two-photon gates were teleported into a quantum circuit with high probability.
Alternatively, using weak cross-Kerr nonlinearities \cite{IHY1985,SZ1997,Boyd1999,LI2000, DingYan2013PLA}, one can induce the requisite interaction between the photons, such as, nearly deterministic linear optical controlled-NOT gate \cite{Nemoto2004}, the deterministic multi-control gates \cite{LinHe2009, LinHeBR2009}, and so on.

In this paper, we focus on the possible experimental realization of the $km$-photon C-GHZ states. We first introduce an entangler of $m$-photon polarization qubits based on weak nonlinearities. Then two schemes are explored: one is creating the $km$-photon C-GHZ state from a $km$-photon GHZ state and the other is combining $k$ $m$-photon GHZ states into the desired state.

\section{Creating the $km$-photon C-GHZ state from a $km$-photon GHZ state}
One of the major ingredient in the present schemes is the entangler of $m$-photon GHZ state introduced in \cite{DingYan2012}. In order to meet the needs of the context, now we describe the mentioned entangler, briefly, as shown in Fig.1.
Consider the initial state $[\frac{1}{\sqrt{2}}(\left|H\right\rangle + \left|V\right\rangle)]^{\otimes m}$ containing $m$ photons ($N_1$, $N_2$, $\cdots$, $N_m$), where the qubits are encoded with the polarization modes $ \left|H\right\rangle \equiv \left|0\right\rangle $ and $\left|V \right\rangle \equiv \left|1 \right\rangle$. Together with the coherent probe beam $|\alpha\rangle$, the whole combined system can be evolved into
\begin{eqnarray}
\left|\psi\right\rangle_\text{ck} & = &[(|HH\cdots H\rangle+|VV\cdots V\rangle)|\alpha\rangle+|HV\cdots V\rangle|\alpha \texttt{e}^{\texttt{i}\theta}\rangle+|VH\cdots H\rangle|\alpha \texttt{e}^{-\texttt{i}\theta}\rangle
\nonumber \\
& & +\cdots +|H\cdots HV\rangle|\alpha \texttt{e}^{(2^{m-1}-1)\texttt{i}\theta}\rangle+|V\cdots VH\rangle|\alpha \texttt{e}^{-(2^{m-1}-1)\texttt{i}\theta}\rangle]/\sqrt{2^m},
\end{eqnarray}
by using linear optical elements and cross-Kerr nonlinear media. The linear optical elements are consisted of polarizing beam splitters (PBSs), which are used to transmit the $|H\rangle$ polarization photons and reflect the $|V\rangle$ polarization photons,
and phase shifter $-(2^{m-1}-1)\theta$. After the following $X$ quadrature homodyne measurement \cite{Nemoto2004,Barrett2005} on the
probe beam and subsequent local operations (a phase shift operation $\phi_i(x)$ corresponding to the value of the homodyne measurement and some selective bit-flip operations $\sigma_x$), according to the classical feed-forward, the initial state can be converted to the $m$-photon GHZ state $\frac{1}{\sqrt{2}} ({\left| {H} \right\rangle}^{\otimes m} + {\left| {V} \right\rangle}^{\otimes m})$.
In addition, if $m$ is odd and the initial state is $(\frac{1}{\sqrt{2}}(\left|H\right\rangle - \left|V\right\rangle))^{\otimes m}$, we noticed that one can obtain the state $\frac{1}{\sqrt{2}} ({\left| {H} \right\rangle}^{\otimes m} - {\left| {V} \right\rangle}^{\otimes m})$, accordingly.

\begin{figure}
  \centering\includegraphics[width=4.5in]{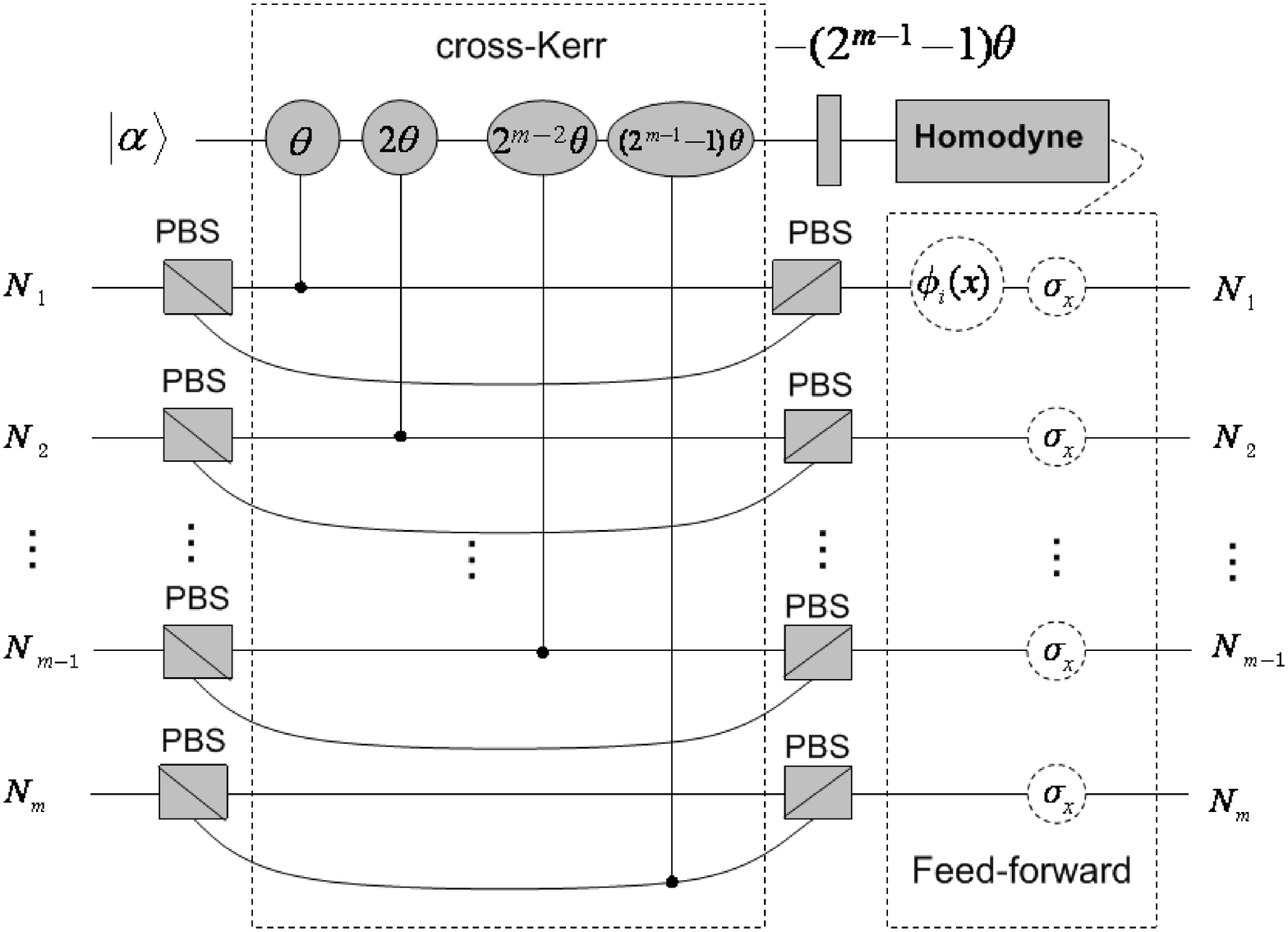}\\
  \caption{An entangler for $m$-photon GHZ state.}\label{m-GHZ}
\end{figure}

\begin{figure}
  \centering\includegraphics[width=2.2in]{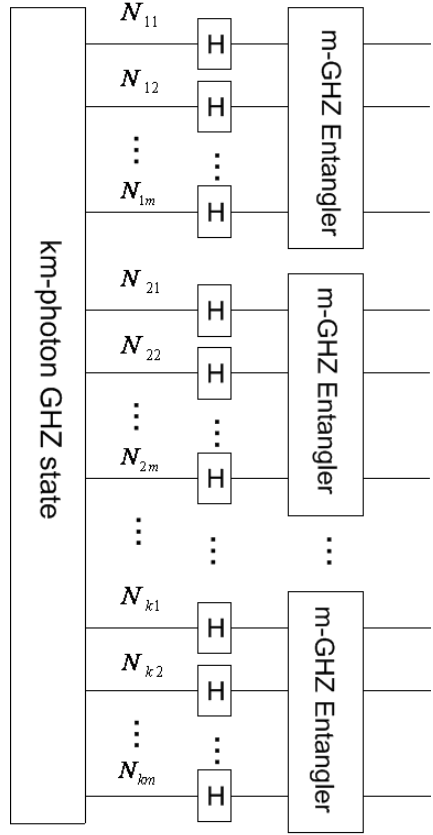}\\
  \caption{The schematic diagram of creating the $km$-photon C-GHZ state from a $km$-photon GHZ state. $km$ photons are divided into $k$ partitions, each partition can be considered as a logical qubit built of $m$ physical qubits.}\label{C-GHZ-1}
\end{figure}

Now we design a setup to generate the $km$-photon C-GHZ state by using the entanglers of $m$-photon GHZ state, where $m$ is odd, as shown in Fig.\ref{C-GHZ-1}. We first suppose that a $km$-photon GHZ state has been generated in the form of $\frac{1}{\sqrt{2}} ({\left| {H} \right\rangle}^{\otimes km} + {\left| {V} \right\rangle}^{\otimes km})$. Later on, one $H$ operation, which evolves the single-photon state $\left| {H} \right\rangle$ (or $\left| {V} \right\rangle$) into the state $\frac{1}{\sqrt{2}} ({\left| {H} \right\rangle}+ {\left| {V} \right\rangle})$ (or $\frac{1}{\sqrt{2}} ({\left| {H} \right\rangle} - {\left| {V} \right\rangle})$) and can be performed by a half-wave plate (the angle between its axis and the horizontal direction is $22.5^{\circ}$), acts on each photon, respectively. As a result, $km$ $H$ operations transform the state $\frac{1}{\sqrt{2}} ({\left| {H} \right\rangle}^{\otimes km} + {\left| {V} \right\rangle}^{\otimes km})$ to $\frac{1}{\sqrt{2}} \{[\frac{1}{\sqrt{2}} (\left|H \right\rangle+ \left|V \right\rangle)]^{\otimes km} + [\frac{1}{\sqrt{2}} (\left|H \right\rangle- \left|V \right\rangle)]^{\otimes km}\}$. At last, we make use of $k$ $m$-photon GHZ entanglers. In the light of the action of the $m$-photon GHZ entangler, obviously, the desired $km$-photon C-GHZ state can be created under the condition that $m$ is odd, as we mentioned above.

\section{Combining $k$ $m$-photon GHZ states into $km$-photon C-GHZ state}

\begin{figure}
  \centering\includegraphics[width=3.5in]{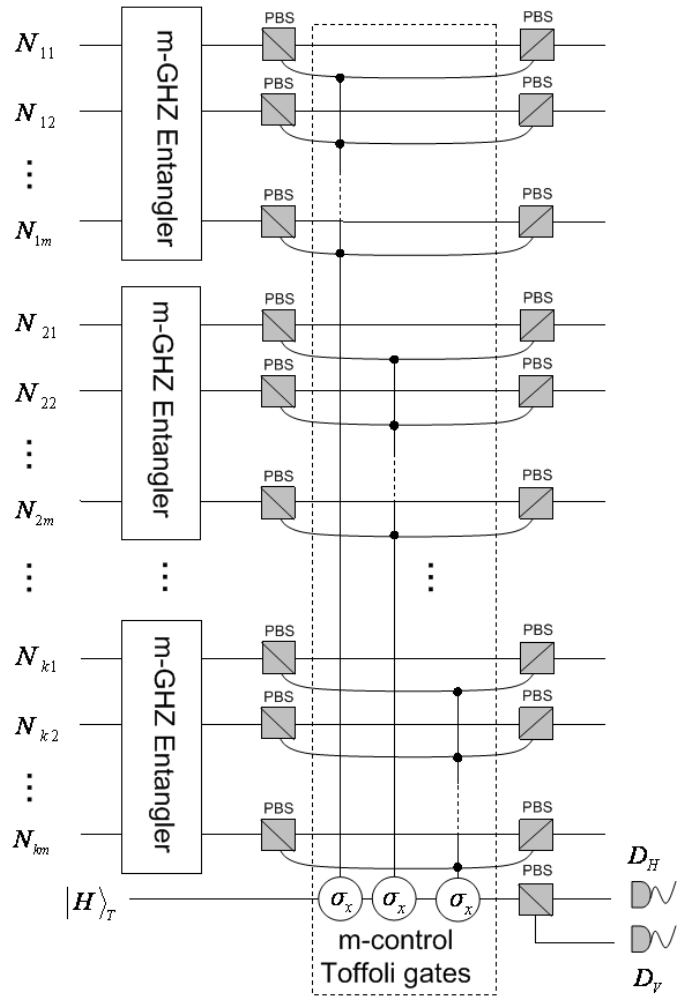}\\
  \caption{The schematic diagram of preparation of $km$-photon C-GHZ states based on $m$-control Toffoli gates.}\label{C-GHZ-2}
  \end{figure}

Generally, we can also create $km$-photon C-GHZ states (as shown in Fig.\ref{C-GHZ-2}) from $km$ single-photon states $[\frac{1}{\sqrt{2}}(\left|H\right\rangle + \left|V\right\rangle)]^{\otimes km}$ by using $k$ entanglers of $m$-photon GHZ state \cite{DingYan2012} and $k$ $m$-control Toffoli gates \cite{LinHe2009, LinHeBR2009}. In order to simplify the following discussion, let us now rewrite two $km$-photon C-GHZ states as
\begin{equation}\label{}
|\phi_{\texttt{C}}^{+}\rangle=\frac{1}{\sqrt{2}}({\left|\texttt{GHZ}_{m}^{+}\right\rangle}^{\otimes k}+{\left|\texttt{GHZ}_{m}^{-}\right\rangle}^{\otimes k})=\frac{1}{\sqrt{2^{k+1}}} [({\left| {H} \right\rangle}^{\otimes m} + {\left| {V} \right\rangle}^{\otimes m})^{\otimes k}
+({\left| {H} \right\rangle}^{\otimes m} - {\left| {V} \right\rangle}^{\otimes m})^{\otimes k}]\equiv|\phi_{\texttt{even}}\rangle,
\end{equation}
\begin{equation}\label{}
|\phi_{\texttt{C}}^{-}\rangle=\frac{1}{\sqrt{2}}({\left|\texttt{GHZ}_{m}^{+}\right\rangle}^{\otimes k}-{\left|\texttt{GHZ}_{m}^{-}\right\rangle}^{\otimes k})=\frac{1}{\sqrt{2^{k+1}}} [({\left| {H} \right\rangle}^{\otimes m} + {\left| {V} \right\rangle}^{\otimes m})^{\otimes k}
-({\left| {H} \right\rangle}^{\otimes m} - {\left| {V} \right\rangle}^{\otimes m})^{\otimes k}]\equiv|\phi_{\texttt{odd}}\rangle,
\end{equation}
where $|\phi_{\texttt{even}}\rangle$ is the $km$-photon C-GHZ state which only contains even (or null) number
of components of $|V\rangle^{\otimes m}$ for each basis of product states, while $|\phi_{\texttt{odd}}\rangle$ is one with only odd
$|V\rangle^{\otimes m}$.

More specifically, we first divide the input state into $k$ units, each consists of $m$ single-photon product state $[\frac{1}{\sqrt{2}}(\left|H\right\rangle + \left|V\right\rangle)]^{\otimes m}$.
Followed by $k$ entanglers of $m$-photon GHZ state, respectively, one can now create $k$ $m$-photon GHZ states $ [\frac{1}{\sqrt{2}} (\left|H \right\rangle^{\otimes m}+ \left|V \right\rangle ^{\otimes m})]^{\otimes k}$. Then, we introduce an auxiliary single-photon state $|H\rangle_{T}$ as the target photon to construct $k$ $m$-control Toffoli gates. The $m$-control Toffoli gate performs a NOT gate operation ($\sigma_x$) on target photon $|H\rangle_T$, controlled by $m$-control photons, i.e., $[(|H\rangle_{1} \langle H|+|V\rangle_{1}\langle V|)\otimes(|H\rangle_{2}\langle H|+|V\rangle_{2}\langle V|)\otimes \cdots \otimes(|H\rangle_{m}\langle H|+|V\rangle_{m}\langle V|)-|VV\cdots V\rangle_{12\cdots m}\langle VV\cdots V|]\otimes (|H\rangle_{T}\langle H|+ |V\rangle_{T}\langle V|)+ |VV\cdots V\rangle_{12\cdots m}\langle VV\cdots V| \otimes (|H\rangle_{T}\langle V|+ |V\rangle_{T}\langle H|)$.
After the actions of $m$-control Toffoli gates, at last, a single-photon detection is made on the auxiliary photon in the \{$|H\rangle, |V\rangle$\} basis.
The single-photon detection can project the combined system onto two expected subspaces, i.e., the states $|\phi_{\texttt{even}}\rangle$ and $|\phi_{\texttt{odd}}\rangle$ with equal probability. In other words, corresponding to the click of $D_H$ or $D_V$, the state $|\phi_{\texttt{C}}^{+}\rangle$ or  $|\phi_{\texttt{C}}^{-}\rangle$ is prepared, alternatively.

\section{discussion and summary}
We design two setups to create the desired $km$-photon C-GHZ states from $km$-photon GHZ state and $km$ single-photon product state respectively. A group of entanglers of $m$-photon GHZ state are introduced in the two schemes, where the cross-Kerr nonlinear media play an important role in experimental realization. In the regime of weak nonlinearity, $(2^{m-1}-1)\theta \simeq 10^{-2}$, there is a difficulty for larger value $m$ since with the increase of $m$ the entangler requires exponentially many nonlinear interactions. If we consider a block size of $m=10$ (see Ref. \cite{FrowisDur2011}), however, the present schemes are feasible with current techniques \cite{LI2000}. In addition, in our second scheme we use $m$-control Toffoli gates which generally requires $O(m^2)$ two-qubit gates \cite{NC2000,BBCD1995}. With the control photons being $m$-photon GHZ state, in present scheme we directly utilize the existing $m$-control Toffoli gate \cite{LinHe2009, LinHeBR2009} since it has been shown that the deterministic optical Toffoli gate only uses the resources increasing linearly with the size of an input.
In summary, we develop two methods for preparing a class of multipartite entangled states---the $km$-photon C-GHZ states, which remain superior stability under the influence of decoherence so that distinct quantum effects might be observed at macroscopic scale.

This work was supported by the National Natural Science Foundation
of China under Grant No: 10971247, Hebei Natural Science Foundation
of China under Grant Nos: A2012205013.

\end{document}